\documentclass[preprint,onecolumn,showpacs,amsmath,amssymb]{revtex4}
\usepackage{times}
\usepackage{graphicx}
\usepackage{dcolumn}
\usepackage{bm}
\usepackage{amstext}

\begin{document}

\title{The role of the elemental nature of A=3 nuclei in neutron rich nuclei}
 
\author{Anisul A. Usmani$^{1}$, Syed Afsar Abbas$^{2}$, Usuf Rahaman$^{1}$, M. Ikram$^{1}$, Farooq Hussain Bhat$^{3}$}

\affiliation{$^1$ Department of Physics, Aligarh Muslim University, Aligarh-202002, India\\ 
$^2$ Centre for Theoretical Physics, JMI University, New Delhi-110 025, India\\
$^3$ Physics Department, Islamic University of Science and Technology,
 Srinagar-192 122, India.}

%


\begin{abstract}
The idea of treating the trinucleon systems as elementary entities in the 
elementary particle model (EPM) as an Effective Field Theory has been a 
success in explaining the weak charge-changing processes in nuclei.
The EPM results are found to be as good as those obtained from nuclear 
microscopic models using two- and three-body forces. We extend this concept 
to investigate the validity of the elemental nature of $A=3$ nuclei through 
studies of nuclear structure of neutron-rich nuclei. By treating neutron-rich 
nuclei as primarily made up of tritons as its building blocks, we extract 
one- and two-triton separation energies of these nuclei. Calculations 
have been performed here within relativistic mean field (RMF) models 
with latest interactions. Clear evidence arises of a new shell structure with 
well-defined predictions of new magic nuclei. These unique predictions
have been consolidated by standard one- and two-neutron separation
energy calculations. The binding energy per nucleon plots of these nuclei
also confirm these predictions.
We make unambiguos prediction of six magic nuclei:
$_{\:\:8}^{24}{\rm O}_{16}$, $_{20}^{60}{\rm Ca}_{40}$,
$_{\:\:35}^{105}{\rm Br}_{70}$, $_{\:\:41}^{123}{\rm Nb}_{82}$,
$_{\:\:63}^{189}{\rm Eu}_{126}$ and $_{\:\:92}^{276}{\rm U}_{184}$. 
\end{abstract}

\maketitle

\section{Introduction}

The Elementary Particle Model (EPM),~\cite{Kammel2010,Measday2001,Kim1965,Congleton1993,Mintz1979,Mintz1973} first 
developed by Kim and 
Primakoff~\cite{Kim1965} is a phenomenological approach, which parametrizes 
the nuclear charge-changing currents in terms of the trinucleon form factors 
in analogy with the corresponding nucleon weak currents. Here, the pair 
$(^3{\rm He} , ^3{\rm H}) \simeq ({h,t})$ is treated as elementary, 
and this is found to give as good results as those obtained with much more 
complicated composite structures for the ground state 
in nuclear microscopic models~\cite{Kammel2010,Measday2001,Kim1965,Congleton1993}. 
We carry this concept forward by studying the role of the elemental nature 
of ($h,t$) in nucleus in general, and in neutron-rich nuclei in particular,
in order to explore some of the pertinent questions that arise in our mind: 
(i) Can triton be understood as a building block for the structures of neutron 
rich nuclei?  
(ii) Are there effective symmetries at $N=2Z$?
(iii) Are there new magicities and new properties
supported by experimental and theoretical evidences which may demand new 
shell structure?  
We, therefore, study the $N=2Z$ neutron-rich nuclei and try to find evidences
for triton clustering based on binding energies. 
First of all, we explore available experimental binding energies of such 
nuclei and extract one- and two-triton separation energies.
Supporting our model, a clear evidence of a magic nucleus, 
$_{\:\:8}^{24}{\rm O}_{16}$ ($N=2Z=16$) appears from the data. 
This one is also supported by the direct 
experimental evidence of a magic number, $N=16$, near the neutron drip 
line~\cite{Ozawa2000}. This may well be interpreted as $N=2Z=16$, a bound 
nuclear system of 8 tritons.
Unfortunately, the data, as in atomic mass compilations~\cite{Pfeiffer2014,Wang2017} 
for $N=2Z$ neutron-rich nuclei, are not available beyond 
$_{17}^{51}{\rm Cl}_{34}$ (17 triton bound system). 

We employ very successful relativistic mean field (RMF) models 
with latest interactions like NL3$^*$~\cite{Ring2009}, NL3~\cite{Ring1997} 
and TM1~\cite{Toki1994} to calculate binding energies of nuclei with $N=2Z$ 
for $Z$ ranging from 5 to 120. We then extract one- and two-triton separation 
energies for a wide spectrum of neutron-rich nuclei and explore for new 
magicities and evidences for new shell structures. 
The experimental data exhibit even-odd effects, and so do the RMF results in
this region. However, for the nuclei with large neutron numbers,
those odd triton number nuclei which carry magic neutron numbers 
become magic.  For example, $_{\:\:35}^{105}{\rm Br}_{70}$, 
$_{\:\:41}^{123}{\rm Nb}_{82}$ and $_{\:\:63}^{189}{\rm Eu}_{126}$. Here, $N=70$
is a harmonic oscillator magic number.  
We predict six prominent magic nuclei:
$_{\:\:8}^{24}{\rm O}_{16}$, $_{20}^{60}{\rm Ca}_{40}$,
$_{\:\:35}^{105}{\rm Br}_{70}$, $_{\:\:41}^{123}{\rm Nb}_{82}$,
$_{\:\:63}^{189}{\rm Eu}_{126}$ and $_{\:\:92}^{276}{\rm U}_{184}$. 
Note that in $_{20}^{60}{\rm Ca}_{40}$, both the numbers 20 and 40, are each
magic in nature. Hence, it is heartening to point out that indeed this nucleus 
has been found to be magic in other mean field calculations~\cite{Terasaki2006,Meng2002}.
Though, in those calculations, the shell gap at harmonic oscillator magic 
number $N=40$ is not as pronounced as at $N=28$ or 50. 
We also obtain, in standard conventional manner, one- and 
two-neutron separation energies for the isotopes of these newly identified 
magic nuclei in order to understand the role played by neutron and proton 
magic numbers, and to investigate if these magicities are being 
translated into triton magic numbers for $N=2Z$ nuclei. 
The standard binding energy per particle plot for all the $N=2Z\le240$ nuclei 
too predicts the same magic nuclei as obtained by extracting
one- and two-triton separation energies. 
The structural properties of these magic nuclei and nearby isotopes have been
studied to understand the staggering.
We present several interesting results, like the six new magic nuclei as stated above. However, given the
significance of superheavy nuclei in current research, this new prediction of a superheavy nucleus 
$_{\:\:92}^{276}{\rm U}_{184}$ stands out.

We discuss the EPM model and triton clustering in the next section, 
followed by the RMF formalism. Thereafter, we present our results with detailed 
discussion. Finally, we present a general conclusion.

\section{The EPM Model and Triton clustering}

The phenomenological EPM,
~\cite{Kammel2010,Measday2001,Kim1965,Congleton1993,Mintz1979,Mintz1973} treats the pair 
$(^3{\rm He} , ^3{\rm H}) \simeq ({h,t})$ as elementary.
In analogy with the corresponding nucleon weak currents, EPM parametrizes 
the nuclear charge-changing currents in terms of the trinucleon form factors.
Amazingly this is found to give as good a result as those obtained with more 
complicated composite structures for the ground state 
in nuclear microscopic models~\cite{Kammel2010,Measday2001,Kim1965,Congleton1993}. 

For example, EPM has been successful in understanding  $\mu^-$ weak capture on 
$^3{\rm He}$, $\mu^- + ^3{\rm He} \rightarrow ^3{\rm H} + \nu_\mu$.
It matches the experimental results as well as the more 
elaborate and extensive microscopic calculations where full nuclear wave 
function which arise from realistic two- and three-body interaction 
terms~\cite{Kammel2010,Measday2001,Kim1965,Congleton1993} are used. 
Using EPM, Mintz~\cite{Mintz1979,Mintz1973} studied the reaction,
$\mu^- + ^6{\rm Li} \rightarrow ^3{\rm H} +  ^3{\rm H} + \nu_\mu$.
Taking clue from the Glashow-Salam-Weinberg model, one writes the matrix
element for the above process as,

\begin{eqnarray}
&&<^3{\rm H}(1),^3{\rm H}(2),\nu|{\rm H}_{\rm W}^{(0)}|^6{\rm Li},\mu^{-}>
\nonumber\\
&=&\frac{G\cos\theta_{c}}{\sqrt{2}}\bar{u}_{\nu}(1-\gamma_{5})
u_{\mu }<^3{\rm H}(1),^3{\rm H}(2)|J_{\lambda}^{\dagger}(0)|^6{\rm Li}>,
\end{eqnarray}
where $\cos \theta_{c}$ = 0.98 with $\theta_{c}$ is the Cabbibo angle and the 
weak coupling constant G = 1.02 $\times {10}^{-5} m_{p}^{-2}$ and
$J_{\lambda}(0) = V_{\lambda}(0) - A_{\lambda}(0)$.
Here, V and A are the vector and axial vector part of the hadronic weak current.
Next, one draws a parallel between the reactions
$\mu^- + ^6{\rm Li} \rightarrow ^3{\rm H} +  ^3{\rm H} + \nu_\mu$ and 
$\mu^- + d \rightarrow n + n + \nu_\mu$.  The current matrix elements
$< n n |   {\rm A}_{\lambda}^{\dagger}(0) | d  >$ and $ < n n |  
{\rm V} _{\lambda}^{\dagger}(0) | d  >$ thus have the same structure as
$< ^3{\rm H} ^3{\rm H} |   {\rm A}_{\lambda}^{\dagger}(0) | ^6{\rm Li}  >$  
and  $< ^3{\rm H} ^3{\rm H} |   {\rm V}_{\lambda}^{\dagger}(0) | ^6{\rm Li} >$,
respectively. These are needed to evaluate the above matrix element in Eq.~(1).

Relevant physically measurable quantities are determined in terms of four form 
factors which are obtained from data from reactions 
$\gamma + ^6{\rm Li} \rightarrow ^3{\rm H} +  ^3{\rm He}$ , 
$^3{\rm H} +  ^3{\rm He}  \rightarrow ^6{\rm Li} + \gamma$ and
$\pi^- + ^6{\rm Li} \rightarrow ^3{\rm H} +  ^3{\rm H}$
by using CVC and PCAC~\cite{Mintz1973}. This model is very successful in fitting the data,
thus confirming the validity of the EPM. 

Phenomenologically, various Effective Field Theoretical (EFT) models, 
motivated by or based on QCD are known. These have acquired more acceptability 
and hence more respectability~\cite{Dobado1997} in recent years. 
Appelquist-Cerazzone theorem determines whether a particular EFT would be 
renormalizable or not. However, the non-renormalizable EFT are no less basic. 
Essentially, the parameters of the EFT Lagrangian carry in them 
information of the underlying more fundamental field theory. It is for this 
reason that even the non-renormalizable EFT's have become important 
phenomenological tool in theoretical physics. Hence, we emphasize the basic 
significance of the EPM model as a rich and useful EFT of the Standard Model. 

We further seek for more evidences from other studies in nuclear physics, 
which treat the ($h,t$) pair as elementary. Are there any? Well, indeed there 
are! Within the sphere of low energy nuclear structure studies a new group 
${\rm SU_{\mathcal{A}}(2)}$ called nusospin has been proposed~\cite{Abbas2005,Abbas2004,Abbas2001,Abbas2016}. 
Just as one takes the pair ($p,n$) as forming 
the fundamental representation of the nuclear SU(2) isospin group, in the 
same manner one hypothesizes that the pair ($h,t$) forms the fundamental 
representation of the new nusopin ${\rm SU_{\mathcal{A}}}(2)$ group. 
The physical justification of this new model are also discussed in 
detail~\cite{Abbas2005,Abbas2004,Abbas2001,Abbas2016}. In support of the 
nusopin group, we have found strong empirical evidences favouring $A=3$ clustering 
in nuclei~\cite{Abbas2011}.  So the EPM model in particle physics, 
finds unequivocal support from the nuclear structure successes of the 
nuclear ${\rm SU_{\mathcal{A}}}(2)$ nusospin group.
Both of these justify the treatment of the pair (h,t) as a fundamental entity.

Now, having provided justification for the nusospin group 
${\rm SU_{\mathcal{A}}}(2)$, here for the sake of completeness  
and clarification of these points, we would like to emphasize a few 
simple supporting points and evidences. We hope that this will act as a simple reminder
of what we already know about $A=3$ clustering in nuclei.

First, say as to what is the justification of treating $^{24}$O as being made 
up of eight tritons?  We argue as  follows. 
Just as light $N=Z$ nuclei with $A=4n$, $n=1,2,3,4...$ may be treated as 
being composed of $n$-$\alpha$ cluster,~\cite{hodgson1994,merchant1994} 
in Table 1, we show several neutron-rich nuclei which may be treated 
as being composed of $n$-clusters of  $^3_1 \rm H_2$. 
We write the binding energy of these nuclei as 
\begin{equation}
E_B = 8.48 n + C k, 
\end{equation}
where 8.48 MeV is the binding energy of $^3_1\rm H_2$. 
We take these $n$-cluster of tritons as forming $k$ bonds and with $C$ as 
inter-triton-bond energy.
We are assuming here the same geometric structure of clusters in 
these nuclei as conventionally done for $\alpha$-clusters in 
$A=4n$ nuclei~\cite{Freer2007}. So all numbers arise from similar configurations.
Thus, the model seems to hold out well with inter-triton cluster bond energy 
of about 5.3 MeV. However, note that the $C$ value of $^{21}$N is somewhat on the 
lower side, but the same was true
of the corresponding alpha cluster $^{28}$Si nucleus with respect to the other 
$\alpha$-cluster nuclei (see Fig. 4 of Ref.~\cite{Freer2007}).

\begin{table}[ht]
\caption{Inter-triton cluster bond energy of neutron-rich nuclei.}
\renewcommand{\tabcolsep}{0.40cm}
\renewcommand{\arraystretch}{0.7}
\begin{tabular}{ccccc}
\hline\hline
Nucleus& $n$& $k$& ${\rm E}_{\rm B}$-8.48n(MeV) &$C$(MeV)\\
\hline
$^{9}{\rm Li}$ &3&3 &19.90 &6.63\\
$^{12}{\rm Be}$&4&6 &34.73 &5.79\\
$^{15}{\rm B}$ &5&9 &45.79 &5.09\\
$^{18}{\rm C}$ &6&12&64.78 &5.40\\
$^{21}{\rm N}$ &7&16&79.43 &4.96\\
$^{24}{\rm O}$ &8&19&100.64&5.30\\
\hline\hline
\end{tabular}
\label{tab1}
\end{table}

As we stated in 2001, Ref.~\cite{Abbas2001}, ``we notice that the value seems 
to work for even heavier neutron-rich nuclei. 
For example for $_{14}^{42}{\rm Si}_{28}$ the inter-triton cluster 
energy is still 5.4 MeV.''  Note that in our model $_{14}^{42}{\rm Si}_{28}$ 
is taken to be made up of an even number of 14-tritons.  
Note the important fact that this nucleus has a large 12-neutron excess over 
the heaviest stable silicon nucleus. In the triton picture, this  nucleus
is found to be extra stable compared to the neighbouring $N=2Z$ nuclei 
empirically as well as theoretically though it does not appear as a well
defined magic number. In complete variance with theoretical model 
predictions,~\cite{warner1994,warner1996,terasaki1997,lalazissis1998,vretenar1999,peru2000,guzman2002}
Fridmann et. al.~\cite{Fridmann2005} showed that $_{14}^{42}{\rm Si}_{28}$ 
was a spherical and a highly magic nucleus. Later, a collapse of $N=28$ shell 
closure in  $_{14}^{42}{\rm Si}_{28}$ was reported by 
Bastin et. al.~\cite{Bastin2007}. They found it to be a well deformed oblate rotor,
but its extra binding still indicated a magic character 
~\cite{Pfeiffer2014,Wang2017}. That is what we notice here in the 
triton picture. 

Next, we point out a strong experimental evidence of the possible 
existence of helion and triton clusters in $^6$Li nuclei. 
Indeed, the same has been very convincingly demonstrated through direct 
trinucleon knockout - both triton and helion, from $^6$Li via exclusive 
electron reaction~\cite{connelly1998}. 
Mirror reactions $^6\rm Li(e,e'^3\rm He)^3\rm H$ and 
$^6\rm Li(e,e'^3\rm H)^3\rm He$ were measured.
The momentum transfer dependence was found to be in complete disagreement 
with the fundamental spectrum of a direct-single nucleon knockout. 
On the other hand, the momentum dependence was in good agreement with a 
direct $A=3$ knockout mechanism. 
This clearly demonstrated that $h$- and $t$-clusters existed as primary 
entities in $^6$Li.

The $N \sim Z$ nuclei are very well explained by the 
SU(2) isospin group with ($p,n$) pair providing basis for a description for 
these nuclei~\cite{Moeller1995}.
As the nuclear SU(2) isospin generates and validates the shell model 
structure of $N \sim Z$ nuclei, we extrapolate this logic to find a 
suitable shell model structure generated by and validated by
the ${\rm SU_{\mathcal{A}}}(2)$ nusospin group. We have already seen how the 
nusospin group predicts new magic numbers of the ($Z,N$) pair of (4,8), 
(6,12), (8,16), (10,20)~\cite{Abbas2005}.
So it is logical to assume that there may be a different shell 
structure associated with this new group.

One-proton and one-neutron (as well as two-proton and two-neutron) 
separation energies play an important role in determining magic 
numbers~\cite{gambhir1990,lalazissis1996} and 
we have pursued similar ideas~\cite{Abbas2005}. 
This is mostly within the framework where SU(2) symmetry based on 
elementarity of the ($p,n$) pair is basic. We extrapolate this idea to the 
${\rm SU_{\mathcal{A}}}(2)$ nusopin with the ($h,t$) pair forming elemental 
entities. Thus, we should be able to talk of one- and two-triton 
separation energies in neutron-rich nuclei while treating these as made up of 
tritons as elementary entities.

\section{The RMF Formalism}

The relativistic mean field theory has been successful in reproducing the 
experimental observations throughout the periodic table, near as well as far 
from the stability line~\cite{gambhir1990,serot1986,boguta1977,serot1992,ring1996,horowitz1981,price1987,crp1991,pannert1987,Vretenar2005,Meng2006,Meng2013,Liang2015,Meng2015,Zhou2016}. 
It has also been pursued to examine cluster 
structures inside the nuclei~\cite{sharma2006,patra2007,Ebran2012,Lu2014}. 
The relativistic Lagrangian for a
many-body system contains all the information of nucleon-nucleon interaction, 
via exchanges of $\sigma$-, $\omega$- and $\rho$-mesons. It is 
written as~\cite{gambhir1990,ring1996,horowitz1981,price1987,crp1991,pannert1987}
\begin{eqnarray}
{\cal L}&=&\bar{\psi_{i}}\{i\gamma^{\mu}
\partial_{\mu}-M\}\psi_{i}
+{\frac12}(\partial^{\mu}\sigma\partial_{\mu}\sigma
-m_{\sigma}^{2}\sigma^{2})	\nonumber\\		   
&-&{\frac13}g_{2}\sigma^{3}
-{\frac14}g_{3}\sigma^{4}	
-g_{s}\bar{\psi_{i}}\psi_{i}\sigma 		
-{\frac14}\Omega^{\mu\nu}\Omega_{\mu\nu} \nonumber \\     
&+&{\frac12}m_{w}^{2}V^{\mu}V_{\mu}		
-g_{w}\bar\psi_{i}\gamma^{\mu}\psi_{i}V_{\mu}    
+{\frac12}c_{4}(V_\mu V^\mu)^2 \nonumber \\
&-&{\frac14}B^{\mu\nu}B_{\mu\nu} 
+{\frac12}m_{\rho}^{2}{\vec{R}^{\mu}}{\vec{R}_{\mu}} 
-g_{\rho}\bar\psi_{i}\gamma^{\mu}\vec{\tau}\psi_{i}\vec{R^{\mu}} \nonumber \\
&-&{\frac14}F^{\mu\nu}F_{\mu\nu}                    
-e\bar\psi_{i}\gamma^{\mu}\frac{\left(1-\tau_{3i}\right)}{2}\psi_{i}A_{\mu} \;,   
\end{eqnarray}
where $\psi$ represents Dirac spinors for nucleons with mass M.
The quantities $m_{\sigma}$, $m_{\omega}$, $m_{\rho}$ are the 
masses assigned to $\sigma$-, $\omega$-, $\rho$-mesons, respectively. 
The $\sigma$, $V_{\mu}$ and $R_{\mu}$ are the fields of ${\sigma}$-meson, 
${\omega}$-meson and ${\rho}$-meson, respectively. 
The quantities $g_s$, $g_{\omega}$, $g_{\rho}$ and $e^2/4{\pi}=$1/137 
are the coupling constants for ${\sigma}$-, ${\omega}$-, ${\rho}$-mesons 
and photon fields, respectively. The $g_2$ and $g_3$ are the self-interaction 
coupling constants for the ${\sigma}$-mesons. 
The quantity $c_4$ is the self-interaction coupling constant 
for ${\omega}$-meson, and this term is used in the TM1 potential. 
The field tensors of the vector, isovector mesons and and of 
the electromagnetic field are given by
\begin{eqnarray} 
\Omega^{\mu\nu}& =& \partial^{\mu} V^{\nu} - \partial^{\nu} V^{\mu} \;,\nonumber \\
B^{\mu\nu}& =& \partial^{\mu}R^{\nu} - \partial^{\nu}R^{\mu}\;,  \nonumber \\
F^{\mu\nu}& =& \partial^{\mu}A^{\nu} - \partial^{\nu}A^{\mu} \;.
\end{eqnarray}
The classical variational principle is used to solve the 
field equations for bosons and fermions.
The Dirac equation for the nucleon is written as:
\begin{equation} 
[-i\alpha.\nabla +\beta (M+S(r))+V(r)]\psi_i=\epsilon_i\psi_i. 
\end{equation}
Here, V(r) and S(r) represent the vector and scalar potential is, defined as 
\begin{equation} 
V(r)=g_{\omega}V_{0}(r)+g_{\rho}\tau_{3}R_{0}(r)+e\frac{(1-\tau_3)}{2}A_0(r)\; ,
\end{equation}
and
\begin{equation} 
S(r)=g_{\sigma}\sigma(r) \;,
\end{equation}
where subscript $i$ stands for neutron(n) and proton(p), respectively.
The field equations for bosons are
\begin{eqnarray} 
\{-\bigtriangleup+m^2_\sigma\}\sigma(r) &=& -g_\sigma\rho_s(r)-g_2\sigma^2(r)-g_3\sigma^3(r)\;,   \nonumber \\ 
\{-\bigtriangleup+m^2_\omega\}V_0(r) &=& g_\omega\rho_v(r)+c_4 V_0^3(r)\; ,          \nonumber \\
\{-\bigtriangleup+m^2_\rho\}R^0_3(r) &=& g_\rho\rho_3(r) \;,             \nonumber \\
-\bigtriangleup A_0(r) &=& e\rho_c(r)\; .				\nonumber 
\end{eqnarray}
Here $\rho_s$, and $\rho_v$ are 
the scalar and vector density for $\sigma$- and $\omega$-fields 
in nuclear system which are expressed as
\begin{eqnarray} 
\rho_s(r) &=& \sum_ {i=n,p}\bar\psi_i(r)\psi_i(r)\;,                         \nonumber \\
\rho_v(r) &=& \sum_{i=n,p}\psi^\dag_i(r)\psi_i(r) \;.			      
\end{eqnarray}
The vector density $\rho_3(r)$ for $\rho$-field and charge density 
$\rho_c(r)$ are expressed by 
\begin{eqnarray} 
\rho_3(r) &=& \sum_{i=n,p} \psi^\dag_i(r)\gamma^0\tau_{3i}\psi_i(r)\; ,		      \nonumber \\
\rho_c(r) &=& \sum_{i=n,p} \psi^\dag_i(r)\gamma^0\frac{(1-\tau_{3i})}{2}\psi_i(r)\;.    
\end{eqnarray}
The quadrupole deformation parameter $\beta_2$ is extracted 
from the calculated quadrupole moments of neutrons and protons through 
\begin{equation}
Q=Q_n+Q_p=\sqrt{\frac{16\pi}{5}}(\frac{3}{4\pi}AR^2\beta_2),
\end{equation}
where $R=1.2A^{1/3}$ fm.\\
The various rms radii are defined as
\begin{eqnarray}
\langle r_p^2\rangle &=& \frac{1}{Z}\int r_p^{2}d^{3}r\rho_p\;,        \nonumber \\
\langle r_n^2\rangle &=& \frac{1}{N}\int r_n^{2}d^{3}r\rho_n\;,        \nonumber \\
\langle r_m^2\rangle &=& \frac{1}{A}\int r_m^{2}d^{3}r\rho\;,          
\end{eqnarray}
for proton, neutron and matter rms radii, respectively.
The quantities $\rho_p$, $\rho_n$ and $\rho$ are their 
corresponding densities. 
The charge rms radius can be found from the proton rms radius 
using the relation $r_{c} = \sqrt{r_p^2+0.64}$ taking into 
consideration the finite size of the proton.
The total energy of the system is obtained by 
\begin{eqnarray}
E_{total} &=& E_{part}(N)+E_{\sigma}+E_{\omega}+E_{\rho}+E_{c}+E_{pair}+E_{c.m.},   
\end{eqnarray}
where $E_{part}(N)$ is the sum of the single-particle energies 
of the nucleons.
$E_{\sigma}$, $E_{\omega}$, $E_{\rho}$, $E_{c}$, 
$E_{pair}$ and $E_{cm}$ are the contributions of meson fields, 
Coulomb field, pairing energy and the center-of-mass energy, respectively.

It is worth mentioning that pairing correlations play an 
important role for open shell nuclei. 
It has significant impact on binding mechanism as well as shape of 
the nuclei~\cite{hansen1987,gambhir1990,patra1993}.
There are three possibilities of pairing for example, 
neutron-proton, proton-proton and neutron-neutron correlation.
We account the pairing into consideration in a way similar 
to Refs.~\cite{gambhir1990,patra1993}.
To take care of pairing effect, the constant 
gap BCS approximation is used throughout the calculations.
In case of simple BCS prescription, the 
expression of pairing energy is written by 
\begin{equation} 
E_{pair}=-G\bigg[\sum_{i\;>0}u_iv_i\bigg]^2\;,
\end{equation}
where G is the pairing force constant, and $v_i^2$ and 
$u_i^2=1-v_i^2$ are the occupation probabilities.
The variation with respect to the occupation numbers, $v_i^2$, 
is expressed by the well-known BCS equation
\begin{equation} 
2\epsilon_iu_iv_i-\bigtriangleup(u_i^2-v_i^2)=0\;,
\end{equation}
with $\bigtriangleup=G\sum_{i>0}u_iv_i$.
The occupation number $n_i$ is given by
\begin{equation} 
n_i = v_i^2 =\frac{\displaystyle1}{\displaystyle2}\Bigg[1-\frac{\displaystyle \epsilon_i-
\lambda}{\displaystyle \sqrt{(\epsilon_i-\lambda)^2+\bigtriangleup^2}}\Bigg]\; ,
\end{equation}
where $\epsilon$ is the single-particle energy for the state $i$. 
The chemical potential $\lambda$ for protons (neutrons) is obtained requiring 
\begin{equation} 
\sum_{i} n_i= Z(N).
\end{equation}
The sum is taken over proton (neutron) states. 
The value of constant gap (pairing gap) for proton and neutron are 
determined from the phenomenological formula of Madland 
and Nix~\cite{madland1988} which are given as 
\begin{equation} 
\bigtriangleup_n = \frac{r}{N^{1/3}}exp(-sI-tI^2)\;, 
\end{equation}
and
\begin{equation} 
\bigtriangleup_p = \frac{r}{Z^{1/3}}exp(-sI-tI^2) \;, 
\end{equation}
where $I=(N-Z)/A$, $r=5.73 MeV$, $s=0.117$, and $t=7.96$.
In particular, for the solution of the RMF equations with 
pairing, we never calculate the pairing force constant G explicitly. 
But the occupation probabilities are directly calculated using the gap 
parameters ($\bigtriangleup_n$ and $\bigtriangleup_p$) 
and the chemical potentials ($\lambda_n$ and $\lambda_p$) for 
neutrons and protons, whereas chemical potentials are determined 
by the particle numbers for protons and neutrons. 
And now, the expression of pairing energy is simplified to
\begin{equation} 
E_{pair}=-\bigtriangleup\sum_{i\;>0}u_iv_i \;.
\end{equation}
The centre-of-mass correction is included by non relativistic 
expression .i.e $E_{c.m.}=-\frac{3}{4}41A^{-1/3}$ MeV.

The RMF calculations are simplified by taking the various symmetries into 
consideration, like conservation of parity, time-reversal symmetry and 
no-sea approximation which eliminates all spatial components of the meson 
fields and Dirac anti-particle contribution of the physical 
observables~\cite{gambhir1990}. 
Moreover, it is not an easy task to compute the binding energy and quadrupole 
moment of odd-$N$ or odd-$Z$ or both odd-$N$ and odd-$Z$ systems. 
In RMF calculations with the effect of time-reversal symmetry, the 
spatial components of vector fields are eliminated which are 
odd under time-reversal and parity. However, spatial components of vector 
fields play an important role in determination of magnetic 
moment~\cite{hofmann1988} but these have very little impact on the bulk 
properties of the nucleus such as binding energy, quadrupole deformation 
and radii~\cite{lalazissis1999}. We pursued our calculation in this context. 
For dealing with odd-$Z$ nuclei, we employ the Pauli blocking 
approximation, which restores the time-reversal symmetry and as a result 
reveals the even-odd staggering very nicely~\cite{Patra2001,Kumar2015}.
A pair of nucleons with spin up/down or spin down/up has a mirror image and 
therefore time reversal symmetry is obeyed. But in case of odd-$A$ or odd-odd 
nuclei, lone odd nucleon fills a quantum state with spin up but its 
corresponding conjugate spin down state remains empty, which 
violates the time reversal symmetry. To take care of this, Pauli blocking
approximation is used. First, we carry out free calculations without any 
blocking and obtain the level of maximum occupancy of the lone odd nucleon 
for minimum energy configuration and then block it in that level either with 
spin up or down. In this approach, the odd nucleon stays in one of  
the  conjugate states, $\pm$m, which is taken out from the pairing scheme. 
Thus, the nucleon of this state is not allowed to fluctuate to other levels. 
The rest of the system has even number of nucleons, which obeys time reversal 
symmetry. This blocking scheme known as Pauli blocking almost restores the time
reversal symmetry. But, it doubles our effort as we need to 
perform our calculations twice.  

\section{Results and Discussion}

A nucleus defined as bound state of $Z$ number of tritons may be written as
${\rm ^{A=3Z}_{\:\:\:\:\:\:\:\:Z} X_{N=2Z}} \equiv {\rm ^A_ZX_N}$.
First of all, we look for the available experimental binding energies for
such nuclei and extract one- and two-triton separation energies, which 
may respectively be obtained as,
\begin{center}
${\rm s_{1t}}={\rm B}({\rm ^A_ZX_N})
-{\rm B}({\rm _{Z-1}^{A-3} Y_{N-2}})-{\rm B}(_1^3{\rm H}_2)$ and\\
${\rm s_{2t}}={\rm B}({\rm ^A_ZX_N})
-{\rm B}({\rm ^{A-6}_{Z-2} Y_{N-4}})-2{\rm B}(_1^3{\rm H}_2)$
\end{center}
where, ${\rm B}({\rm ^A_ZX_N})$ is the binding energy of the nucleus
${\rm ^A_ZX_N}$. The experimental binding energies are not available beyond 
${\rm N}_t=17$ bound systems. We, therefore, resort to RMF theory to extend 
our calculation for a wide spectrum of $N=2Z$ nuclei, $5\le {\rm N}_t\le120$.
The RMF binding energies as well as ${\rm s_{1t}}$ and ${\rm s_{2t}}$ agree 
with experimental data to a great extent in their overlap region.  We plot 
${\rm s_{1t}}$ and ${\rm s_{2t}}$ as a function of triton numbers 
(${\rm N_t}$) in Fig.~1. 

\begin{center}
\begin{figure}
\vspace{0.30cm}
s\includegraphics[scale=0.53]{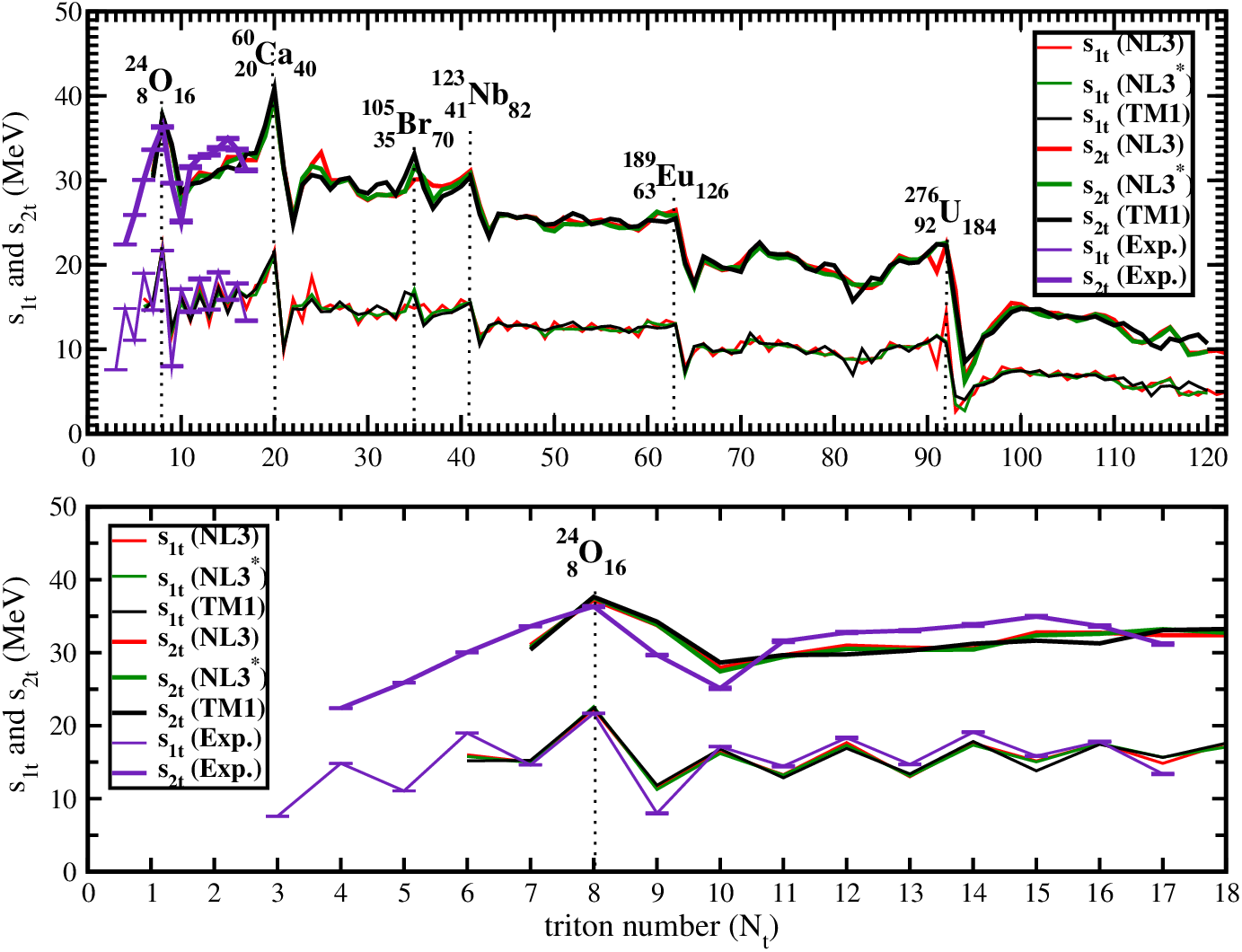}
\caption{One- and two-triton separation energies. In the lower panel,
the region of experimental data has been emphasized.}
\label{fig1}
\end{figure}
\end{center}

The most prominent feature in Fig.~1 is the first peak shown by data and 
RMF both e.g., for ${\rm N_t}$$=$8 {\it i.e.} for $^{24}_{\:\:8}{\rm O}_{16}$ 
and an equally sharp dip for ${\rm N_t}$$=$9 i.e., for $^{27}_{\:\:9}{\rm F}_{18}$.
We know that such drops in one-neutron and one-proton separation energies
when going from one $Z/N$ number to the next one is a signal of the magicity 
character of a particular $Z/N$ number. In the context of our discussion 
here, magicity means a much stronger binding for a particular number
of tritons as compared to the adjoining number of tritons.
Hence, ${\rm N_t}$$=$8 is a magic number with respect to different 
bound states of tritons. 
Besides limited experimental data the vast RMF results clearly show 
${\rm N_t}$$=$8, 20, 35, 41, 63 and 92 as magic numbers, which correspond to 
$N=$16, 40, 70, 82, 126 and 184. The experimentally observed magic numbers
are 8, 20, 50, 82 and 126. 
The magic number 184 is  predicted to be the next neutron magic number after 
126 by many models~\cite{Kruppa2000,Bender1999,Sil2004,Li2014} yet to be 
determined experimentally. 
In Ref.~\cite{Li2014}, the magicity at $N=184$ is associated with $Z=120$; that is
$_{120}^{304}X_{184}$. The $N=184$ magicity in our work is  however true only for
$Z=92$. This $N=2Z$ link is basic in our work.

\begin{figure}
\vspace{0.30cm}
\includegraphics[scale=0.50]{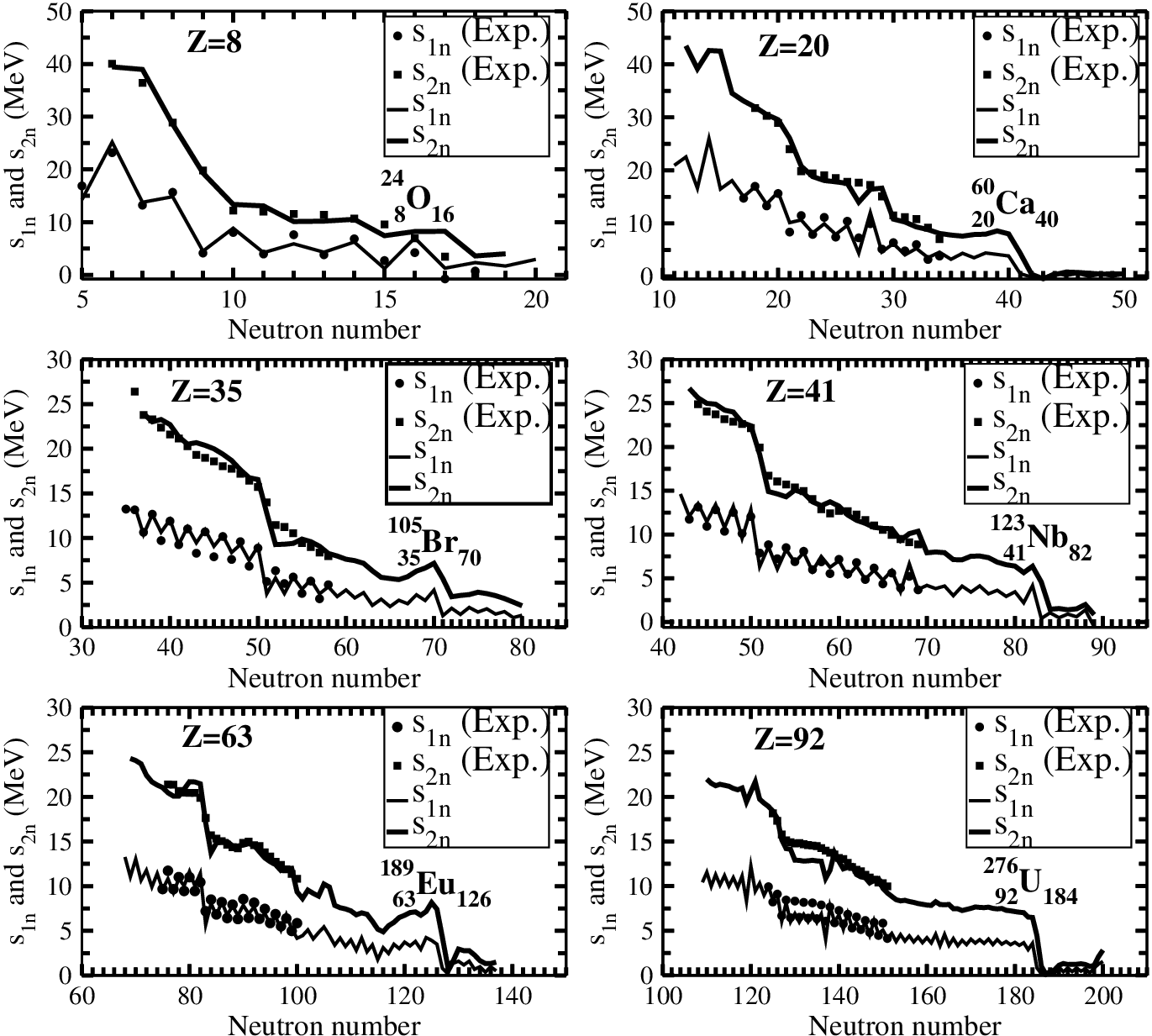}
\caption{One- and two-neutron separation energies 
      (${\rm s}_{1n}$ and ${\rm s}_{2n}$, respectively) for the isotopes of 
      the newly identified magic nuclei in Fig-1.}
\label{fig2}
\end{figure}

\begin{table}
\caption{Structural properties of $N=2Z$ exotic and surrounding nuclei}
\renewcommand{\tabcolsep}{0.40cm}
\renewcommand{\arraystretch}{0.7}
\begin{tabular}{ccccccc}
\hline\hline
Nuclei& BE(MeV)& $\beta_2$& $r_c$& $r_p$& $r_n$& $r_m$\\
\hline
$^{22}$O&163.230&	0.00580&2.738&	2.618&	3.086&	2.924\\
$^{23}$O&167.052&	0.00505&2.739&	2.620&	3.163&	2.985\\
$^{24}$O&171.448&	0.00468&2.748&	2.629&	3.246&	3.054\\
$^{25}$O&172.711&	0.04971&2.773&	2.655&	3.325&	3.126\\
$^{26}$O&175.033&	0.00553&2.798&	2.681&	3.394&	3.192\\
$^{58}$Ca&458.972&	0.00280&3.610&	3.520&	4.084&	3.898\\
$^{59}$Ca&463.127&	0.00199&3.626&	3.537&	4.115&	3.929\\
$^{60}$Ca&466.997&	0.00180&3.642&	3.553&	4.148&	3.960\\
$^{61}$Ca&467.520&	0.00301&3.652&	3.564&	4.186&	3.993\\
$^{62}$Ca&467.389&	0.00504&3.663&	3.575&	4.219&	4.022\\
$^{103}$Br&805.876&	0.04394&4.345&	4.271&	4.860&	4.668\\
$^{104}$Br&807.213&	0.07139&4.356&	4.282&	4.881&	4.688\\
$^{105}$Br&813.038&	0.01404&4.359&	4.285&	4.911&	4.712\\
$^{106}$Br&814.330&	0.03821&4.370&	4.297&	4.930&	4.730\\
$^{107}$Br&816.457&	0.03909&4.380&	4.306&	4.951&	4.750\\
$^{121}$Nb&946.281&	0.04916&4.558&	4.487&	5.029&	4.852\\
$^{122}$Nb&950.256&	0.00105&4.562&	4.491&	5.043&	4.865\\
$^{123}$Nb&952.640&	0.01638&4.571&	4.501&	5.060&	4.881\\
$^{124}$Nb&953.040&	0.04968&4.577&	4.507&	5.098&	4.910\\
$^{125}$Nb&954.094&	0.04661&4.583&	4.512&	5.130&	4.936\\
$^{187}$Eu&1408.843&	0.06467&5.293&	5.232&	5.775&	5.598\\
$^{188}$Eu&1412.772&    0.00389&5.293&	5.232&	5.782&	5.604\\
$^{189}$Eu&1416.262&    0.00405&5.300&	5.239&	5.796&	5.617\\
$^{190}$Eu&1416.738&    0.00242&5.305&	5.245&	5.822&	5.637\\
$^{191}$Eu&1417.305&	0.06915&5.320&	5.260&	5.850&	5.662\\
$^{274}$U&1945.325&     0.00007&6.035&	5.982&	6.530&	6.351\\
$^{275}$U&1948.208&	0.00387&6.039&	5.986&	6.544&	6.363\\
$^{276}$U&1951.805&	0.00005&6.042&	5.989&	6.559&	6.374\\
$^{277}$U&1952.650&	0.00020&6.049&	5.996&	6.575&	6.389\\
$^{278}$U&1952.829&	0.00024&6.059&	6.006&	6.590&	6.403\\
\hline\hline
\end{tabular}
\label{shape}
\end{table}

The magic nuclei that appear in the triton picture are 
$_{\:\:8}^{24}{\rm O}_{16}$, $_{20}^{60}{\rm Ca}_{40}$,
$_{\:\:35}^{105}{\rm Br}_{70}$, $_{\:\:41}^{123}{\rm Nb}_{82}$,
$_{\:\:63}^{189}{\rm Eu}_{126}$ and $_{\:\:92}^{276}{\rm U}_{184}$.
Are these magic nuclei an effective manifestation of proton and neutron magic
numbers? We investigate it in Fig.~2, wherein we plot one- and two-neutron 
separation energies (${\rm s_{1n}}$ and ${\rm s_{2n}}$) for the isotopes of 
these nuclei.  This figure shows that for $N=$ 40, 70, 82, 126 and 184
there is a sharp fall indicating magicities at $Z=N/2$ i.e., triton numbers 
${\rm N}_t$=20, 35, 41, 63 and 92. A less significant fall is
also seen for $N=50$ in the plots of $Z=35$ and $Z=41$ isotopes (middle row),
which do not show up as magic numbers in the triton picture. 
We also  note that $_{\:\:8}^{24}{\rm O}_{16}$ is a magic nucleus, which is 
in line with the findings of the study~\cite{Ozawa2000}. 
However, no staggering is seen 
for $N=16$ in top left panel of Fig.~2. We conclude that for the heavier nuclei 
with large number of neutrons, it is neutrons which decide the effective 
magicities in terms of tritons, and for lighter nucleus like 
$^{24}_{\:\:8}{\rm O}_{16}$, it is protons which play this role. 
Probably, due to this reason $N=70$ for $_{\:\:35}^{105}{\rm Br}_{70}$, though
less significant, it still turns out to be a magic number. The $N=40$ too appears to be a 
magic number in Fig.~1, which is confirmed by the staggering
seen at $N=40$ in case of $_{20}^{60}{\rm Ca}_{40}$ in the top right panel of
Fig.~2.  But for this nucleus, $Z=20$ too is an experimentally observed magic 
value, and we notice the highest peak for it in Fig.~1.  
The result is in line with the previous studies~\cite{Terasaki2006,Meng2002}, 
wherein $^{60}_{20}{\rm Ca}_{40}$ is found to be a doubly magic nucleus 
though the shell gap at $N=40$ is not as pronounced as at $N=28$ and 50.  
Interestingly, $N=40$ and 70 are magic values for the harmonic oscillator 
potential.

\begin{figure}
\vspace{0.30cm}
\includegraphics[scale=0.50]{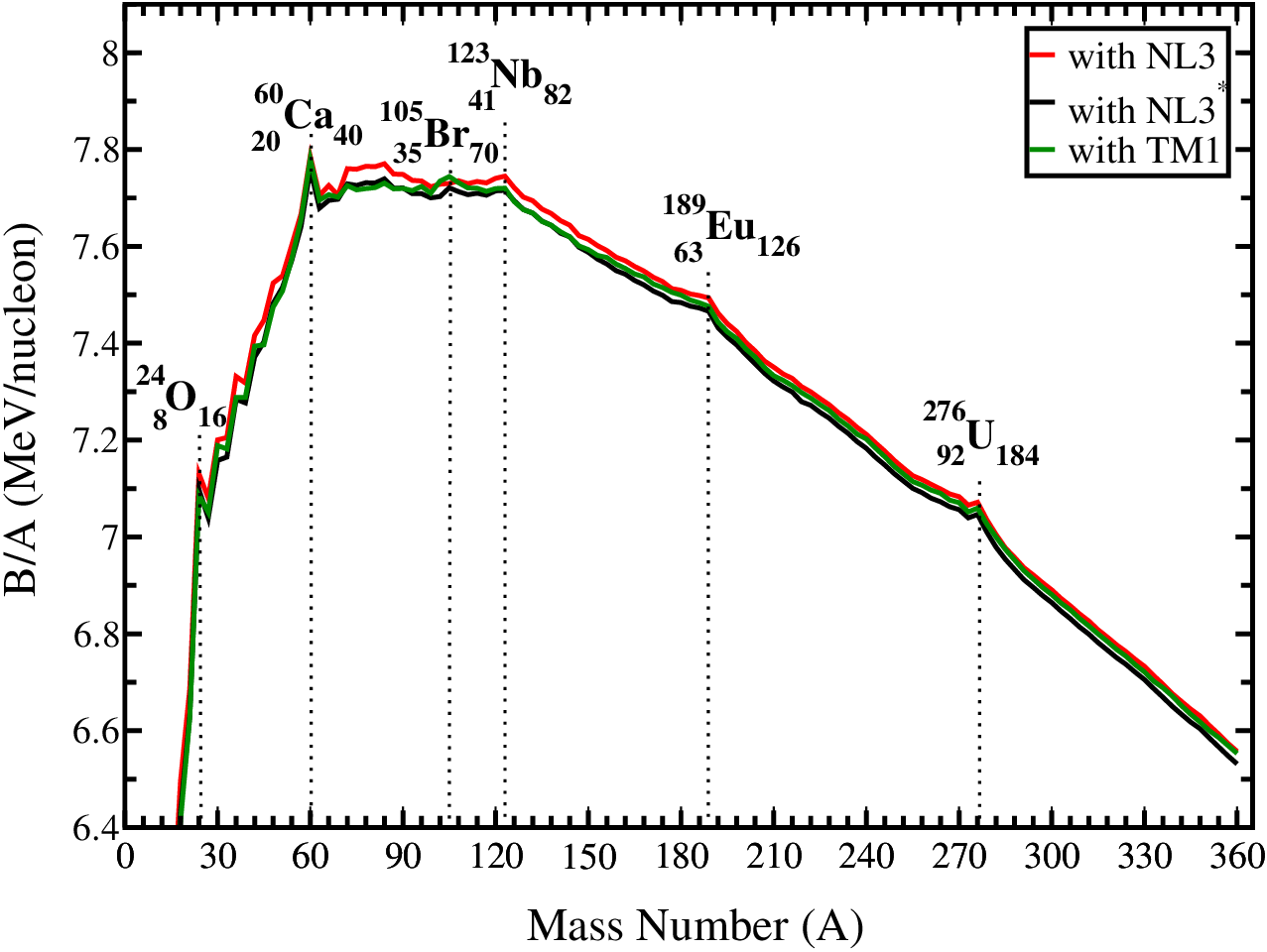}
\caption{B/A for $_{\:\:\:\:\:\:\:\:Z}^{A=3Z}{\rm X}_{N=2Z}$ nuclei}
\label{fig3}
\end{figure}

We find that whenever triton number is even, the triton separation energy is 
significantly higher than the adjoining odd triton numbers except for those 
odd triton numbers whose neutron number, $N=2Z$, is magic.  The nuclei, 
$_{\:\:35}^{105}{\rm Br}_{70}$, $_{\:\:41}^{123}{\rm Nb}_{82}$,
and $_{\:\:63}^{189}{\rm Eu}_{126}$ are more bound with respect to their
neighbouring even triton number nuclei because of the magicities of the 
neutrons they carry.
Thus, we observe an effective manifestation of nucleon odd-even effect and 
that the nucleon magicities are being translated to those of tritons.

We further investigate the structural properties of these magic nuclei and 
calculate their quadrupole deformations.  We observe smaller departure from 
sphericity for these newly identified magic nuclei compared 
to their nearby isotopes as given by deformation parameter ($\beta_2$) 
in Table~2. The nucleus gets more deformed and less bound with an extra triton
which appears to go in the next higher shell. 
The staggering at these magic nuclei may be attributed to it.
The radii are found to increase with increasing mass number. 
Our study shows that an effective shell structure of the bound 
states of tritons seems to  be manifesting here.

In Fig.~3, we plot binding energy per nucleon for all the 
$_{\:\:\:\:\:\:\:\:Z}^{A=3Z}{\rm X}_{N=2Z}$ nuclei studied here. 
It too shows clear magicities of  the same set of
nuclei: $_{\:\:8}^{24}{\rm O}_{16}$, $_{20}^{60}{\rm Ca}_{40}$,
$_{\:\:35}^{105}{\rm Br}_{70}$, $_{\:\:41}^{123}{\rm Nb}_{82}$,
$_{\:\:63}^{189}{\rm Eu}_{126}$ and $_{\:\:92}^{276}{\rm U}_{184}$.
This consolidates our assertions above on new magicities.
The above new neutron-rich magic nuclei and in particular the superheavy 
$_{\:\:92}^{276}{\rm U}_{184}$ nucleus, are the most unique predictions of our 
model here.  

\section{Conclusions}

What seems to be happening in the neutron-rich nuclei is that the degree
of freedom appears to be changing from ($p,n$) of ${\rm SU_{\rm I}}(2)$ isospin
structure to ($h,t$) of ${\rm SU_{\mathcal{A}}}(2)$ nusospin structure, so much
so that for ${\rm ^{3Z}_{\:Z} X_{2Z}}$ nuclei the predominant structure 
is that of $Z$-tritons. As neutron number is always even for the $N=2Z$ nuclei, the
odd triton number (odd proton number) decides the spin and hence the shell model
predictions of spin remain unaltered. The zero spin may be assigned to all even
triton nuclei. This may appear to be amazing to those who wish to continue treating
($p,n$) as being the only degree of freedom relevant for all nuclei: 
($N \sim Z$) nuclei and as well as very neutron-rich nuclei.

Against the theoretical predictions that $N=28$ shell closure will be destroyed 
and that $_{14}^{42}{\rm Si}_{28}$  will be highly deformed~\cite{warner1994,warner1996,terasaki1997,lalazissis1998,vretenar1999,peru2000,guzman2002},
the empirical evidence~\cite{Fridmann2005} showed that this indeed was 
spherical and a magic nucleus.  However, at variance with it, 
Bastin et al.~\cite{Bastin2007} reported a collapse of $N=28$ shell closure, 
wherein they found it to be a well-deformed oblate rotor. The experimental binding
energies of N=2Z nuclei still predicted it to be extra stable, though not a magic 
nucleus~\cite{Pfeiffer2014,Wang2017}. We observe exactly the same in the 
triton picture in Fig.~1. As no staggering is seen after this nucleus, it can 
not be predicted as being magic. The $_{14}^{42}{\rm Si}_{28}$ has 12-neutron excess 
over the heaviest stable silicon nuclide.
Nature appears to be more enterprising.  As neutron number increases 
${\rm SU_{I}}(2)$ isospin group leads to an induced
${\rm SU_{\mathcal{A}}}(2)$ nusospin group.  

We have already shown in other 
papers~\cite{Abbas2005,Abbas2004,Abbas2001,Abbas2016}
as to how ${\rm SU_{\mathcal{A}}}(2)$ nusospin finds justification.
In Ref.~\cite{Abbas2011}, we have enumerated several strong empirical evidences 
in support of triton-helion cluster structure effects in nuclei.
The triton picture of $N=2Z$ nuclei as well as conventional studies of one- 
and two-neutron seperation energies indicate magicity of some special nuclei, 
e.g.,  $_{\:\:8}^{24}{\rm O}_{16}$, $_{20}^{60}{\rm Ca}_{40}$,
$_{\:\:35}^{105}{\rm Br}_{70}$, $_{\:\:41}^{123}{\rm Nb}_{82}$,
$_{\:\:63}^{189}{\rm Eu}_{126}$ and $_{\:\:92}^{276}{\rm U}_{184}$.
The superheavy nucleus, $_{\:\:92}^{276}{\rm U}_{184}$, may be easily 
accessible to experimental confirmation due to its relatively small charge, $Z=92$.

Thus, what we call nusospin model here and what is called Elementary Particle 
Model, are essentially talking about the same physical reality of ($h,t$) being 
fundamental and elementary, though using them in different framework of 
nuclear studies; the former one in nuclear strong interaction studies, while 
the latter in the electro-weak studies.

With this hindsight, one now looks at Fig.~1, and then the wisdom of the
${\rm SU_{\mathcal{A}}}(2)$ nusospin model dawns upon us with the possibility
of a new shell structure of tritons. The magicity of eight-tritons is already 
confirmed by the empirical study of $_{\:\:8}^{24}{\rm O}_{16}$. The binding
energy per nucleon plot as in Fig.~3, which predicts the same magic nuclei as
in triton picture of Fig.~1, consolidates our assertions.
We, therefore, urge the experimentalists to conduct studies for the possible 
confirmations of these exotic magic nuclei.

\section*{Acknowledgments}
\noindent AAU wishes to thank Inter-University Centre for Astronomy and 
Astrophysics, Pune, India for providing the Visiting Associateship 
under which part of this work was carried out. We thank the anonymous referee
for a detailed report, which has helped us in improving the presentation
considerably.

\end{document}